\let\ifarxiv=\iftrue     
\pdfoutput=1

\ifarxiv

\documentclass[12pt,a4paper]{article}
\usepackage[a4paper,text={450pt,650pt},centering]{geometry}

\fi

\ifarxiv\else

\documentclass[11pt,a4paper]{article}
\usepackage{mathptmx}
\usepackage[a4paper,text={130mm,198mm}]{geometry}

\fi

\usepackage{amsmath,amssymb}
\usepackage[bookmarks=true,hyperfigures=true]{hyperref}
  \usepackage{graphicx}
\usepackage[nosort]{cite}

\usepackage{graphicx}

\let\oldbfseries=\bfseries
\let\oldmdseries=\mdseries
\let\oldnormalfont=\normalfont
\renewcommand{\bfseries}{\oldbfseries\boldmath}
\renewcommand{\mdseries}{\oldmdseries\unboldmath}
\renewcommand{\normalfont}{\oldnormalfont\unboldmath}

\allowdisplaybreaks[3]

\numberwithin{equation}{section}

\usepackage[font=small,labelfont=bf,width=0.85\textwidth]{caption}

\providecommand{\hypersetup}[1]{}
\providecommand{\texorpdfstring}[2]{#1}

\hypersetup{plainpages=false}
\hypersetup{pdfpagemode=UseNone}
\hypersetup{bookmarksnumbered=true}
\hypersetup{pdfstartview=FitH}
\hypersetup{colorlinks=false}
\hypersetup{citebordercolor={.5 1 .5}}
\hypersetup{urlbordercolor={.5 1 1}}
\hypersetup{linkbordercolor={1 .7 .7}}

\providecommand{\arxivref}[2]{\href{http://arxiv.org/abs/#1}{#2}}

\providecommand{\href}[2]{#2}
\providecommand{\arxivlink}[1]{\href{http://arxiv.org/abs/#1}{arxiv:#1}}

\newcommand{\pa}{\partial}
\newcommand{\be}{\begin{equation}}
\newcommand{\ee}{\end{equation}}
\newcommand{\bea}{\begin{eqnarray}}
\newcommand{\eea}{\end{eqnarray}}
\def \ci{\cite}
\def \YY {{\rm Y}}

\newcommand{\no}{\nonumber}
\newcommand{\rf}[1]{(\ref{#1})}

\newcommand{\cn}{\mathop{\mathrm{cn}}\nolimits}

\def \sql {{\sqrt{\l}}\ }
\def \del{\partial}
\def \a {\alpha}

\def\g{\gamma}
\def\s{\sigma}

\def \ov {\over}
\def\la{\label}

\def\J{{\cal J}}
\def \om {\omega}
\def\E{{\cal E}}
\def\w{\omega}
\def\b{\beta}
\def\l{\lambda}

\def \adss{$AdS_5 \times S^5$\ } 

\def \r { \rho}

\def \p {\phi}
\def \vp {\varphi}

\def \ads {{$AdS_5$}}

\def \ha {{1 \over 2}}
\def \la{\label}

\def \k {\kappa}
\def\foot{\footnote}
\def \const {{\rm const}}
\def \NN {{\rm N}} 
\def \J {{\cal J}}

\def\rr {{\rm r}}

\def\appendix#1{
  \addtocounter{section}{1}
  \setcounter{equation}{0}
  \renewcommand{\thesection}{\Alph{section}}
  \section*{Appendix \thesection\protect\indent \parbox[t]{11.15cm}
  {#1} }
  \addcontentsline{toc}{section}{Appendix \thesection\ \ \ #1}
  }
\newcommand{\eq}[1]{(\ref{#1})}

\def\be{\begin{equation}}
\def\ee{\end{equation}}
\def \bi{\bibitem}
\def \la {\label}

\def \ha {{1 \over 2}}
\def \td {\tilde} 
\def \N {{\cal N}}
 
\def \const {{\rm const}}
 
\def\S{{\cal S} }
\def \XX {{\rm X}}

\def \zz {{\rm z}}
\def \n {\nu} 

\def\[{\begin{equation}}
\def\]{\end{equation}}
\def\<{\begin{myeqnarray}}
\def\>{\end{myeqnarray}}

\def \ep {\epsilon}
\def \k {\kappa}
\def \N {{\mathcal N}}

\def \J {{\mathcal J}}

\def \a {\alpha}
\def \E {{\mathcal E}}
\def \b {\beta}
\def \g {\gamma}
\def \G {\Gamma}
\def \d {\delta}
\def \l {\lambda}

\def \m {\mu}
\def \n {\nu}
\def \s {\sigma}
\def \S {\Sigma}
\def \r {\rho}

\def \p {\phi}

\def \vp {\varphi}

\def \frac#1#2{{ #1 \over #2}}
\def \td {\tilde}

\def \adss {$AdS_5 \times S^5\ $}
\def \ads {$AdS_5$\ }

\def \N {{\mathcal N}}

\def \s { \sigma }

\def \vp {\varphi}
\def \fourth {{1 \ov 4}}

  \def \td { \tilde }

\def \la {\label}
\def \om {\omega}
\def \del{\partial}

\def \la {\label}

\def \ha {{1 \over 2}}
\def \ov {\over}

\def \w  {\omega}

\def \sql {{\sqrt{\l}}\ }

\def\E{{\mathcal E}}
\def\w{\omega}
\def\b{\beta}
\def\l{\lambda}

\def \ads {{$AdS_5$}}
\def \s{\sigma}
\def \pa{\partial}

\def \ha {{1 \over 2}}

\def \k {\kappa}
\def\foot{\footnote}
\def \const {{\rm const}}

\def\be{\begin{equation}}
\def\ee{\end{equation}}
\def \bi{\bibitem}

\def \ha {{1 \over 2}}
\def \td {\tilde}
\def \ci{\cite}
\def \N {{\mathcal N}}
 
\def \const {{\rm const}}

\def\S{{\mathcal S} }

\def \XX {{\rm X}}
\newenvironment{myeqnarray}{\arraycolsep0pt\begin{eqnarray}}{\end{eqnarray}\ignorespacesafterend}
\newenvironment{myeqnarray*}{\arraycolsep0pt\begin{eqnarray*}}{\end{eqnarray*}\ignorespacesafterend}

\def \cn {{\rm cn}}
\def \be {\bea} 
\def \ee {\eea}
\def \te {\textstyle}
\def \ba {\bea} 
\def \ea {\eea}  
\def \EE  {{\bf E}}
\def \KK {{\bf K}}

\def \ha {{\textstyle {1 \ov 2}} }

 \def \E {{\rm E}}
 \def \S {{\rm S}}
 \def \J {{\rm J}}

\def \zz {{\rm z} } \def \w {\omega}
\def \rr {{\rm r}}

\begin{document}


\thispagestyle{empty}
\phantomsection
\addcontentsline{toc}{section}{Title}

\begin{flushright}\footnotesize%
\texttt{Imperial-TP-AT-2010-05},
\texttt{\arxivlink{1012.3986}}\\
overview article: \texttt{\arxivlink{1012.3982}}%
\vspace{1em}%
\end{flushright}

\begingroup\parindent0pt
\begingroup\bfseries\ifarxiv\Large\else\LARGE\fi
\hypersetup{pdftitle={Review of AdS/CFT Integrability, Chapter II.1: Classical AdS5xS5 string solutions}}%
Review of AdS/CFT Integrability, Chapter II.1:\\
Classical $AdS_5 \times S^5$ string solutions
\par\endgroup
\vspace{1.5em}
\begingroup\ifarxiv\scshape\else\large\fi%
\hypersetup{pdfauthor={A.A. Tseytlin}}%
A.A.\ Tseytlin\foot{Also at Lebedev Institute, Moscow.}
\par\endgroup
\vspace{1em}
\begingroup\itshape
Blackett Laboratory, Imperial College London, SW7 2AZ, U.K. 
\par\endgroup
\vspace{1em}
\begingroup\ttfamily
tseytlin@imperial.ac.uk
\par\endgroup
\vspace{1.0em}
\endgroup

\begin{center}
\includegraphics[width=5cm]{TitleII1.mps}
\vspace{1.0em}
\end{center}

\paragraph{Abstract:}
We  review basic examples of classical   
string  solutions  in $AdS_5 \times S^5$.
We concentrate on  simplest rigid  closed string  solutions 
of circular or folded type described by integrable 1-d Neumann system 
 but  mention 
also various generalizations and  related open-string solutions.

\ifarxiv\else
\fi
\hypersetup{pdfsubject={MSC (2010): 81T30, 83E30}}%

\ifarxiv\else
\fi
\hypersetup{pdfkeywords={strings in AdS space, solitons, spinning strings, integrable models}}%

\newpage


\section{Introduction}

\adss  space  plays a  special  role in superstring theory \ci{grs}. 
This space (supported by a    5-form  flux)
 is one  of the three maximally supersymmetric ``vacua'' of type IIB 10-d 
supergravity \ci{schw}, along with  its limits -- the flat Minkowski space  and the plane-wave 
background \ci{hul}.  It appears as a ``near-horizon'' region  of the solitonic  D3-brane 
 background \ci{hs}; that  explains its central role in the  AdS/CFT duality \ci{mald}
(see \ci{og} for a   review). The duality states that certain ``observables'' 
in $\N=4$ supersymmetric  $SU(N)$ 4-d gauge theory  have direct counterparts 
in the type IIB superstring theory in \adss  space, and vice versa. 

The type  IIB superstring theory in a curved space  with a 5-form  Ramond-Ramond (RR) 
 background is defined by  the  Green-Schwarz \ci{GS} action  ($T_0 ={1\over 2\pi \alpha'}$)
 \bea \la{a}
 && I= I_{_{\rm B}} + I_{_{\rm F} } \ ,\ \ \ \ \ \ \
   \ \ \ \ I_{_{\rm B}}=
 \ha  T_0  \int d^2 \sigma\
\sqrt{-g} g^{ab} G_{\mu\nu}(x)\del_a x^\mu \del_b x^\nu\ ,       \\
&& 
I_{_{\rm F}} = {i} T_0 \int d^2 \s
( \sqrt{-g} g^{ab}\d^{IJ} - \ep^{ab} s^{IJ} )
\bar \theta^I \r_a D_b \theta^J\  + {O} (\theta^4)  \la{f} \   .  \eea
Here $x^\mu$ ($\m=0, 1, ..., 9$)  are the  bosonic string coordinates, 
$\theta^I$ ($I=1,2$) are two   Majorana-Weyl spinor fields,  
$g_{ab}$ ($a,b=0,1$)  is an independent 2-d  metric,  
 $\r_a$ are  projections of the 10-d Dirac matrices,
$
\r_a \equiv \G_{A} E^{A}_\m \del_a x^\m $, 
$E^{A}_\m$ is the vielbein of the target space metric,
$G_{\m\n} = E^{A }_\m E^{B}_\n \eta _{AB}$. 
$ \ep^{ab}$ is antisymmetric 2-d tensor and $s^{IJ}=$diag$(1,-1)$. 
 $D_a$ is the projection of the
10-d covariant derivative
$D_{ \mu}$. The latter is   given by $D_{ \mu}=\del_{ \mu}
+ \fourth\omega^{AB}_{ \m} \Gamma_{AB}
 - { 1 \ov 8 \cdot 5!}
 \G^{ \m_1...\m_5} \G_{ \m}\  F_{ \m_1... \m_5}$, where 
$\omega^{AB}_\m$ is the Lorentz  connection and
$F_{ \m_1... \m_5}$ is the RR 5-form field.
Here $G_{\mu \nu}$  and $F_{ \m_1... \m_5}$ 
 should be related so  that the 2-d Weyl
  and kappa-symmetry anomalies 
cancel.  

In the case of the \adss  background the 
 explicit form of the superstring  action  can be found using the supercoset 
 construction \ci{mt}.
 The group of super-isometries
  (Killing vectors and Killing spinors or  
 solutions of $D_\mu \epsilon^I=0$) of    this background 
 is $PSU(2,2|4)$, i.e. the same as $\N=4$ super-extension of the 4-d conformal group 
 $SO(2,4)$. Using   that  $AdS_5 = SO(2,4)/SO(1,4)$ and $S^5= SO(6)/SO(5)$ 
 the superstring action can be constructed in terms of the components of 
 $PSU(2,2|4)$  current restricted to the   coset $PSU(2,2|4)/[ SO(1,4) \times  SO(5)]$
 (see  \ci{IIb} for details). 
 
Since the metric of \adss  has  direct product structure,  
the  bosonic part of the action \rf{a} is a sum of the 
actions for the $AdS_5$ and $S^5$ sigma models. 
The two sets of bosons are coupled  through their interaction with the fermions.
 The latter fact  is crucial for the UV finiteness of the 
superstring model \ci{mt,dgt,rtt} (see  also  \ci{IIc}).

Below  we shall consider  classical  bosonic solutions of the \adss 
string action. The study of  classical string solutions  
and their  semiclassical quantization  initiated in \ci{bmn,gkp,ft1,ft2} 
is an important tool for  investigating
 the structure of the 
AdS/CFT correspondence (for reviews see, e.g.,  \ci{t1,t2,pl,t4}).
The  AdS  energy of a closed string solution  expressed in terms of other conserved charges
and string tension 
gives the strong coupling limit of the  scaling dimension 
of the corresponding gauge-theory operator.  
Classical solutions for open strings  ending at the  boundary of $AdS_5$ 
describe the strong coupling   limit of the  associated Wilson loops 
and gluon scattering amplitudes  (see \ci{AM}  and  \ci{IVb,IVc,V3}).

Coset  space sigma models  are known to be classically integrable \ci{pohl,lup}
 and this integrability  extends \ci{bpr} also to the full kappa-invariant \adss 
 superstring action. 
 The integrability  allows one to describe, in particular,    large class
  of (finite gap \ci{novik}) 
 classical string solutions in terms of the associated 
 spectral curve  \ci{kaz,grom1} (see  \ci{IId}). 
  
  This description  is, however,      formal 
  and   obscures somewhat 
  the  physical interpretation of the solutions.  
   It is very useful  to complement it with a 
  study of specific  examples of solutions 
  that can be constructed  directly from the  sigma model equations of motion 
  by   starting with  certain  natural ansatze. 
  This  will be our aim below. 
  
  We shall mostly  concentrate on the simplest spinning ``rigid'' closed string solutions 
  for which the shape  of the string does not change with time
  (extra oscillations increase the energy for given spins).  
  We shall consider several types of solutions  and their  limits 
  that reveal different patterns of dependence  of the energy on the string tension 
  and the spins. This   provides an   important information 
  about the strong `t Hooft coupling limit of the corresponding gauge theory anomalous dimensions
  and thus  aids one in  understanding 
    the underlying description of the string/gauge theory 
  spectrum valid  for  all values of the string tension or `t Hooft coupling.

 \section{Bosonic string  in \texorpdfstring{\adss}{AdS5xS5} }
 
 At the  classical level (with  fermion fields vanishing) 
 the $AdS_5$ and $S^5$ parts of the string action 
  are still effectively  coupled  
 through their interaction with 2-d metric $g_{ab}$.
 If one solves  for $g_{ab}$ one gets a non-linear  Nambu-Goto type  action 
 containing   interactions  between the $AdS_5$ and $S^5$ coordinates. 
 In the conformal gauge 
 $\sqrt {-g} g^{ab} = \eta^{ab}$  the  classical  equations  for the  
 $AdS_5$ and $S^5$   parts are  decoupled, but there is a constraint 
 on  their  initial data from the equation for $g_{ab}$, i.e.  
  that the  2-d stress tensor should vanish (the Virasoro conditions). 
 We shall  study the corresponding solutions below  but let us start with 
 the definition of the $AdS_n$ space  and the explicit form of 
 the \adss  bosonic string action. 

 \subsection{\texorpdfstring{$AdS_5\times S^5$}{AdS5xS5} space}
 
 Just like the $d$-dimensional sphere $S^d$  can be represented as a 
 surface  in $R^{d+1}$
 \be\la{kon}
   X_M X_M  = X_1^2 + ... + X^2_{d+1} =1 \ee
 the $d=n+1$  dimensional anti - de Sitter  space $AdS_d$ 
 can be represented as a hyperboloid (a constant negative curvature quadric) 
 \be \la{konn}
 -\eta_{PQ} Y^P  Y^Q = Y_0^2 - Y_1^2 -...- Y_{n}^2 + Y_{n+1}^2 = 1
 \ee
 in $R^{2,d-1}$ with  the metric 
 \be   ds^2 = \eta_{PQ} dY^P  dY^Q \ , \ \ \ \ \ \ \ \ \ 
 \eta_{PQ}=(-1,+1,..., +1,-1)  \ . \ee
 We set the radius  of the sphere and the hyperboloid to 1.  
 In what follows we will be interested in the case of $d=5$. 
 
 It is often useful  to solve  \rf{konn},\rf{kon} 
 by  choosing  an  explicit parametrization 
of   $Y_P$   and  $X_M$ 
in terms of 5+5   independent ``global'' coordinates
\be
\YY_1&\equiv& Y_1+ iY_2 = \sinh \r \ \cos \theta \  e^{i \phi_1}\
,
 \ \ \ \ \
\YY_2 \equiv 
Y_3 + i Y_4  = \sinh \r \ \sin \theta \  e^{i  \phi_2}\  ,  \no
\\
\YY_0 &\equiv& Y_5 + i Y_0 = \cosh \r \  e^{i t }\ , \ \ \ \ \ \ \
\ \ \ \ \ \ \ \
\XX_3\equiv X_5 + i X_6 =
\cos  \g \ e^{ i \vp_3} \ , \la{relx} \\
\XX_1 &\equiv&  X_1 + i X_2 = \sin   \g \ \cos
\psi \ e^{ i \vp_1} \ , \ \ \ \
\XX_2 \equiv X_3 + i X_4 =  \sin   \g \ \sin \psi \
e^{ i \vp_2} \ . \no \ee
Then the corresponding metrics are 
\be\la{ads}
(ds^2)_{AdS_5}
= d \rho^2 - \cosh^2 \rho \ dt^2 + \sinh^2\rho \ (d \theta^2 +
 \cos^2 \theta \ d \phi^2_1 + \sin^2 \theta \ d\phi_2^2) \ , 
\\
\la{ses}
(ds^2)_{S^5}
= d\gamma^2 + \cos^2\gamma\ d\varphi_3^2 +\sin^2\gamma\
(d\psi^2 +
\cos^2\psi\ d\varphi_1^2+ \sin^2\psi\ d\varphi_2^2) \ , 
\ee
and they are obviously related  by an analytic continuation.


Note that choosing  $\r >0$ and  $ 0 < t \leq 2 \pi$ 
(and standard periodicities for the $S^3$ angles 
$\theta,\phi_1,\phi_2$) already  covers the hyperboloid  once.
Near ``the  center''  $\r=0$ the  $AdS_5$  metric is that of $S^1 \times R^4$ 
while near its boundary $\r\to \infty$ it is  that of $S^1 \times S^3$. 
To  avoid closed time-like curves and to  relate the corresponding theory to 
gauge theory in $R \times S^3$  it is standard to  decompactify the $t$ direction, 
i.e. to assume $ - \infty < t < \infty$. Thus in all discussions of AdS/CFT and 
in what follows by $AdS_5$ we shall understand its universal cover
(in particular, we will ignore the possibility of string winding in global AdS time direction).
In the case of $AdS_2$  plotted as a  hyperboloid in $R^{2,1}$ that corresponds to 
going around  the circular dimension infinite number of times or ``cutting it open''.  
We present  images of $S^2$ and of a  universal cover of $AdS_2$ 
in Figure 1.\foot{We thank  N. Beisert  for sending us these figures.}
Another   useful image  of the universal cover  of the  $AdS_3$  space 
is a body of 2-cylinder with $R_t \times S^1$ as a boundary and $\r$ as a radial coordinate.

\begin{figure}[ht]
\centerline{\includegraphics[scale=0.15]{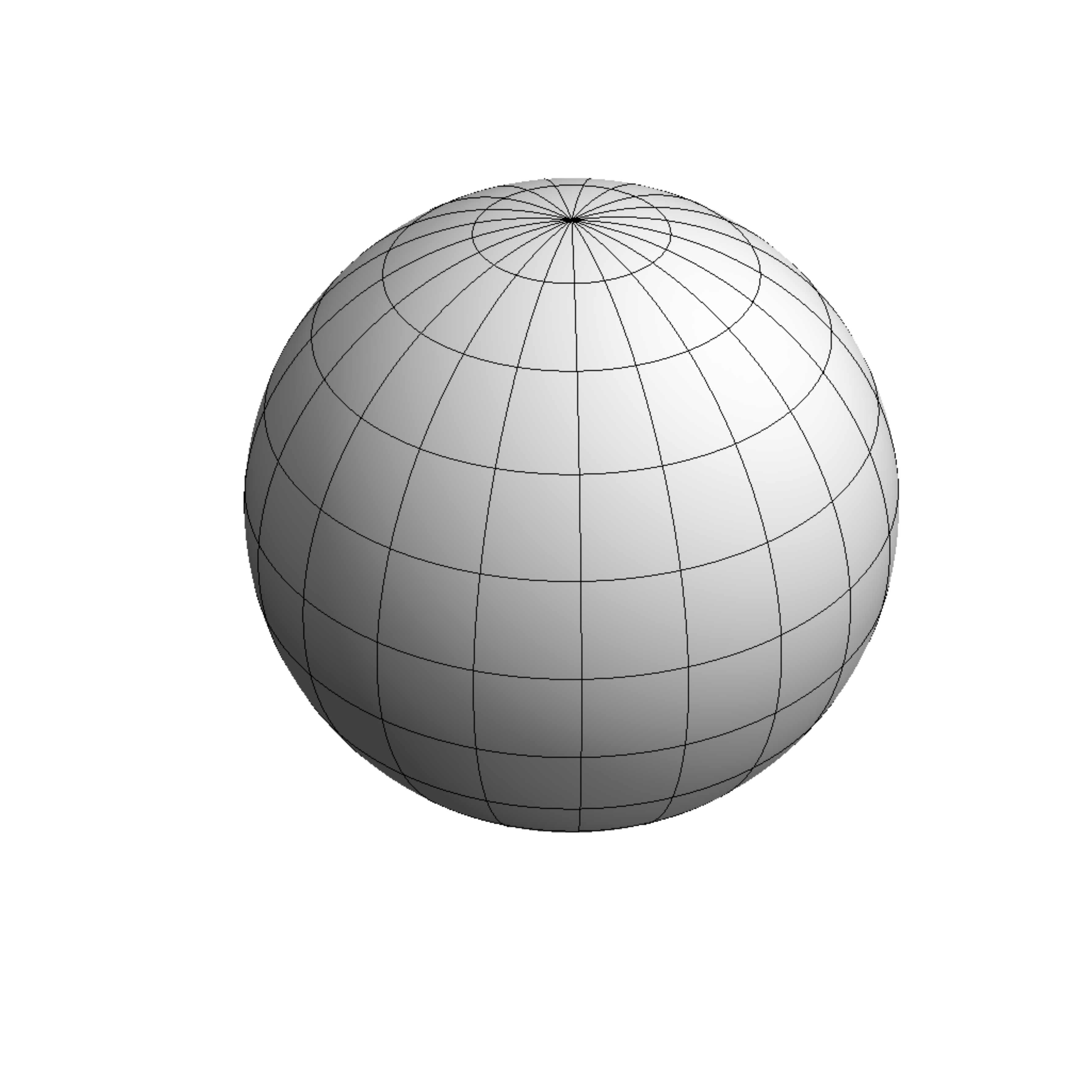}
\includegraphics[scale=0.15]{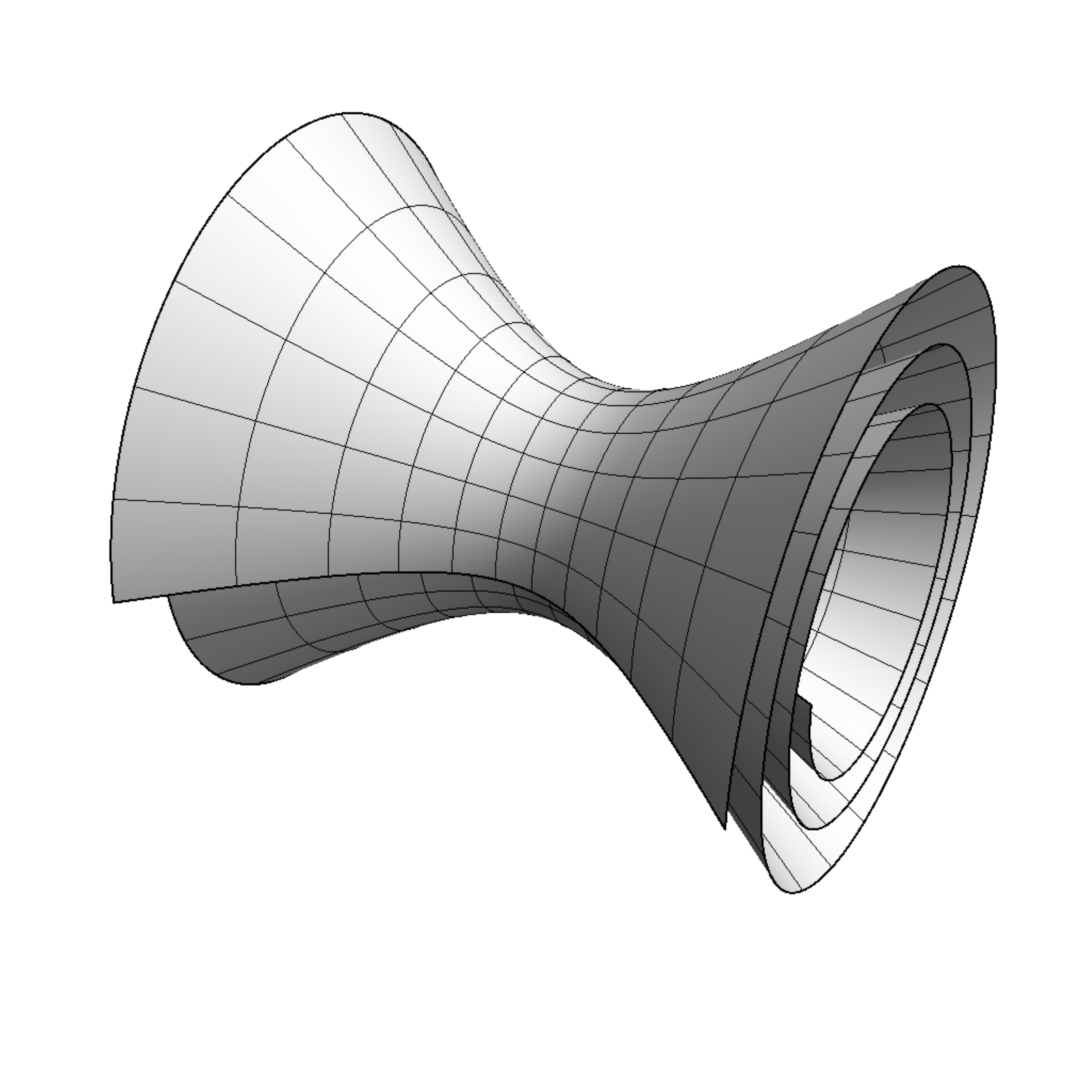} }
\caption{Images of a sphere and of a universal cover of AdS space
\label{sfel}}\nonumber
\end{figure}

 Let us  mention also another choice of \adss  coordinates  -- the Poincar\'e  coordinates -- 
 that cover only part of \ads\  (for more details see, e.g., \ci{og}):
\be
Y_0={x_0\ov z} = \cosh \r \ \sin\ t \ , \ \ \ \ \ 
Y_5= { {1 \ov 2 z}} ( 1 + z^2 - x^2_0 + x_i^2) =  \cosh \r \ \cos\ t \ , \no \\
Y_i= {x_i\ov z} = n_i \sinh \r  \ , \ \ \  
Y_4={  {1 \ov 2 z}} ( -1 + z^2 - x^2_0 + x_i^2) =  n_4 \sinh\r 
  \ , \la{eme}  \ee
Here  $n^2_i + n^2_4 =1$ \ ($i=1,2,3$)
 parametrizes the 3-sphere in \rf{ads}:  \ $dn_k dn_k = d\Omega_3(\theta,\p_1,\p_2)$. 
Then the \ads\ metric \rf{ads} takes the form 
($m,n=0,1,2,3$)
\be\la{aa} 
  (ds^2)_{AdS_5} = {1 \ov z^2} ( dx^m dx_m  + dz^2 )  \ , \ \ \ \ \ \ \ \ \ \ \ 
x_m = \eta_{mn} x^n \ .   \ee 
The full \adss metric may be written also  in the conformally-flat form as 
\be  (ds^2)_{AdS_5 \times S^5} = {1 \ov z^2} ( dx^m dx_m +  dz_{_M}  dz_{_M})  \ , \ \ \ \ \ \  \ 
 z^2 = z_{_M} z_{_M} \ ,  \ \ \ \ \ \ M=1,...,6 \ , \la{zee} \ee 
where $dz_{_M}  dz_{_M}= dz^2 + z^2 d\Omega_5(\gamma,\psi,\vp_1,\vp_2,\vp_3)$. 
The Poincar\'e  coordinates are useful for the discussion of 
solutions representing open  strings ending at the AdS 
boundary  
 (see \ci{IVb,IVc,V3}).

\subsection{String action,  equations of motion and conserved angular momenta}
 
 The bosonic part  of the \adss  action \rf{a}  in the conformal gauge is 
\be
I_{_B}= \ha T \int d\tau \int^{2\pi}_0 {d\sigma} 
 \ (  L_{AdS} + L_S )\ 
,  \ \ \ \ \ \ \ \    T={R^2 \ov 2 \pi \a'} =  { \sql  \ov 2 \pi }\ ,  \la{Lsp} \ee
where $\sql \equiv { R^2 \ov \a'}$  ($\l$ corresponds to `t Hooft coupling 
on the $\cal N$=4 super Yang-Mills side),  \  $R$ is the (same)  radius of 
$AdS_5$ and $S^5$ and 
\bea
\la{SL}
L_{AdS}=- \pa_a Y_P\pa^a Y^P -  
\td \Lambda (Y_P Y^P+1)\, , \  \ \ \ \ \ \ 
L_S=- \pa_a X_M\pa^a X_M+ \Lambda 
(X_M X_M-1)\,  .
\la{SSL}
\eea
Here  $X_M$, $M=1,\ldots , 6$  and $Y_P$, 
 $P=0,\ldots , 5$ are  
the embedding coordinates of ${R}^6$ 
with the Euclidean metric $\delta_{MN}$ in $L_S$ and of $R^{2,4}$ 
 with  $\eta_{PQ}=(-1,+1,+1,+1,+1,-1)$ 
 in $L_{AdS}$,  respectively  ($Y_P = \eta_{PQ} Y^Q$). 
 $\Lambda$  and $\td \Lambda$ are the Lagrange multipliers
 imposing the two  hypersurface conditions. 
 The classical  equations for  \rf{Lsp} 
are 
 \be \la{ade}
&& \del^a \del_a Y_P -  \td \Lambda  Y_ P =0 \ , \ \ \ \ \ \ \ \ \ 
 \td \Lambda =    \del^a  Y_P \del_a Y^P  \ ,\ \ \ \ \ \ 
 Y_P Y^P=- 1\ ,\\
&& \la{sse}
 \del^a \del_a X_M   +  \Lambda  X_M  =0 \ , \ \ \ \ \ \ \ 
 \Lambda  = \del^a X_M  \del_a  X_M  \ , \ \ \ \  X_M X_M =1 \ . \ee  
The action \eq{Lsp} is  to be supplemented 
 with the  conformal gauge  constraints 
 \be \la{cv}
 \dot{Y}_P\dot{Y}^P+  Y'_P  Y'^P
 +   \dot{X}_M\dot{X}_M+ X_M ' X_M ' =0 \ , \ \ \ \ \ \ \
   \dot{Y}_P Y'^P+ \dot X_M  X_M '  =0 \ . \ee
 We will be interested in the closed string solutions with the 
  world sheet  as a  cylinder, i.e. will 
 impose the periodicity conditions 
 \be \la{peri}
 Y_P (\tau, \s + 2\pi) = Y_P (\tau, \s) \ , \ \ \ \ \ \ \ \ 
  X_M (\tau,\s + 2\pi) = X_M (\tau,\s) \ . \ee 
 The action \rf{Lsp} is invariant under the $SO(2,4)$ and $SO(6)$  rotations 
  with the conserved (on-shell) 
 charges 
 \bea \la{cha}
 S_{PQ}= \sqrt{\lambda } \int^{2\pi}_0 {d\s\over 2\pi }
\ (Y_P \dot Y_Q  -  Y_Q\dot Y_P) , \ \ \ \ \ \ 
J_{MN}= \sqrt{\lambda } \int^{2\pi}_0 {d\s\over 2\pi }
\ (X_M\dot  X_N  -  X_N\dot X_M)
\eea
There  is a   natural choice of the 3+3 Cartan generators  of
$SO(2,4) \times SO(6)$ corresponding to the 3+3 linear isometries
of the 
\adss metric \rf{ads},\rf{ses}, 
i.e. to the 
 translations in the  time $t$, in the 2 
angles $\phi_a$   and  the  3 angles  $\vp_i$:
\be \la{aas}
S_0 \equiv  S_{50} \equiv E = \sql \E\ , \ \ \ \  S_1\equiv
S_{12}= 
\sql \S_1 \ , \ \
\  S_2\equiv S_{34} =  \sql \S_2 \ , \\
 \la{jij}
J_1 \equiv  J_{12} = \sql \J_1 \ , \ \ \ \ \ 
J_2 \equiv  J_{34}  =\sql \J_2 \ , \ \ \ \ \ 
J_3 \equiv  J_{56}= \sql \J_3 \ . \ee 

\subsection{Classical solutions: geodesics}

We will be interested  in classical solutions that  have finite
values
of the  AdS energy $E$ and  the spins
$S_r,\ J_i$ ($r=1,2; \  i=1,2,3$). 
 The  Virasoro condition will give a relation  between the 6
 charges in \rf{aas},\rf{jij} allowing one to express the energy in terms 
 of the other 5,  i.e. \ 
$ E= \sql \E(\S_r,\J_i; k_s)=  \sql \E({S_r\ov \sql} ,{J_i\ov \sql}; k_s)$. 
 Here $k_s$   stands for other   (hidden)  conserved   charges, 
 like 
 ``topological'' numbers determining 
 particular shape of the string (e.g., number  of folds, spikes,  winding numbers, 
 etc).\foot{A simple
 example of an infinite-energy solution  is an infinitely stretched string in $AdS_2$ described (in
 conformal gauge) by 
 $t= \k \tau, \ \ \r=\r(\s)$, \ \  $\r'^2 - \k^2 \cosh^2 \r=0$, 
 i.e.  $\cosh \r = |\cos (\k \s )|^{-1} $. 
 It is formally $2\pi$ periodic if $\k=1$. In the Poincare patch the corresponding solution is 
 $z= { \cos \k\s  \ov \cos \k\tau - \sin \k\s  } , \ \ 
 x_0 = { \sin \k\tau  \ov \cos\k \tau - \sin \k\s  } $.}

For a solution to  have a consistent semiclassical interpretation, 
it should  correspond to  a state of a   quantum  Hamiltonian which
carries the same  quantum numbers (and  should thus  be 
 associated to a particular SYM operator with definite scaling dimension).
It  should represent a ``highest-weight'' state of a  symmetry algebra, i.e. 
all other non-Cartan (non-commuting) 
components of the  symmetry generators \rf{cha}  should
vanish;  other members of the multiplet can be obtained by applying rotations 
to a ``highest-weight'' solution.\foot{For a discussion 
of the relation of the above $SO(2,4)$   charges to the standard 
conformal group generators in the boundary theory 
and a relation  between $SO(2,4)$ representations   labelled by the AdS  energy 
$E= S_{50}$ and  the  dilatation operator $D= S_{54}$ see \ci{t2} and refs. there.}

Let us start with  
 point-like strings, for which 
$Y_P=Y_P(\tau), \ X_M=X_M (\tau)$ in  \rf{ade}--\rf{cv}, i.e.  with 
massless   geodesics in \adss.
Then  $\Lambda,\tilde \Lambda=\const$  (as follows directly from  \rf{ade},\rf{sse})
and \rf{cv} implies that $\Lambda = - \td \Lambda >  0$. 
The generic massless geodesic  in \adss  
 can  be of  two ``irreducible'' types
 (up to a global $SO(2,4) \times SO(6)$
transformation): (i) massless 
 geodesic that stays entirely within $AdS_5$; \ \ 
 (ii) a   geodesic 
  that runs along the time direction  in $AdS_5$ 
  and  wraps a big   circle of $S^5$. 
In the latter case the angular motion in $S^5$ provides 
an effective mass to a particle in $AdS_5$, 
i.e. the corresponding geodesic in $AdS_5$ is a massive one,
\be \la{bam}
Y_5+i Y_0 = e^{i \k \tau}\ , \ \
 \ \ \ 
X_5+ i X_6 = e^{i \k \tau} \ , \ \ \ \ \ \k= \sqrt \Lambda  \ , \  \ \ \ 
Y_{1,2,3,4}=X_{1,2,3,4}=0 \ . \ee
The only non-vanishing integrals of motion are 
$E=J_3=\sql  \k$, representing the energy and the 
$SO(6)$ spin  of this BPS state, corresponding to the BMN  ``vacuum'' 
 operator tr$(Z^{J_3})$
 in the SYM theory \ci{bmn} (see also \ci{IIc}). 
 
The solution for a massless 
geodesic in $AdS_5$  is   a straight line in $R^{2,4}$,   
$Y_P(\tau)  = A_P + B_P \tau   $ with $
B_P B^P= A_P B^P=0\ , \ \ A_P A^P=-1 . $
The $SO(2,4)$ angular momentum tensor in  \rf{cha}  is
 $S_{PQ}= \sql (A_P B_Q- A_Q B_P)$. It   always has
non-vanishing non-Cartan components  \ci{t2}, e.g.,  
if $Y_5+ i Y_0 = 1 + i { p }\ \tau , \ \ 
 Y_3= { p } \ \tau  , \ \  Y_{1,2,4}=0 $ 
we get $S_{50}=S_{53}=\sql { p}$.
This geodesic thus does not represent a ``highest-weight''
 semiclassical state.   
In terms of Poincare   coordinates \rf{aa} 
the massless geodesic   is represented by 
$x_0=x_3=p \tau, \ z=a=\const$, i.e.
it runs parallel to the boundary  (reaching  the boundary at
 spatial infinity where Poincare patch ends -- that 
  follows from its description in global coordinates). 

Below  we shall consider examples of  extended ($\s$-dependent) 
solitonic string solutions of the  equations 
 \rf{ade},\rf{sse}  subject to the  constraints \rf{cv},\rf{peri} 
that have finite 
 AdS  energy and spins. 
The  aim will be  to find  the expression for the energy $E$
in terms of other charges.\foot{Early discussions of semiclassical strings 
in de Sitter and Anti de Sitter spaces appeared, e.g., in \ci{barb,deve}.
The fact that in AdS space the string mass  scales linearly  with large quantum numbers 
(as opposed to square root 
Regge relation in flat space) was first observed in \ci{deve}. }
In general, a string  all points  of which can  move fast in $S^5$ 
will admit a ``fast string'' (BMN-type) limit   in which  
 $E$ will have an analytic dependence on the square of 
string tension or on  $\l$   when expressed in terms of 
 $S_r$ and $J_i$  and expanded  in large total spin of $S^5$ \ci{ft1,ft2}.
At the same  time, the energy of a 
 string   whose center is at  rest or which moves 
only within the $AdS_5$ will  depend explicitly on $\sql$ 
(i.e. will be non-analytic in $\l$) \ci{gkp,ft2,ptt2}. 


\section{Simplest    rigid string  solutions}

Here we shall  consider  few simple explicit closed-string  solutions of  the 
non-linear equations \rf{ade},\rf{sse}
which are ``rigid'', i.e. for which  the shape   of the string 
does not change with time. 
These may be interpreted as
 examples of  non-topological solitons  of the \adss conformal-gauge 
string sigma model \rf{Lsp} 
on a  2-d cylinder $(\tau, \s)$.


\subsection{Examples of  string solutions in flat space}

Let us  start with recalling several  examples of   string 
solutions in flat  space. 
The flat-space string action  and equations of motion 
 in the conformal gauge  are ($\sqrt {-g} g^{ab} = \eta^{ab}$, \
$x_\m = \eta_{\m \n} x^\n$, \ $\eta_{\m\n} = {\rm diag}(-1,1, ..., 1) $) 
\bea \la{bf}
  I_{_{\rm B}}=
  \ha T_0  \int d^2 \sigma\ \del_a x_\mu \del^a  x^\mu\ , \ \ \ \ \ \ \   
  \del_+ \del_- x^\m =0 \ , \ \ \ \ \ \ 
\del_\pm   x^\m \del_\pm x_\m  =0   \ .  
\eea
The general solution of   free  equations
$x^\m = x^\m_0 +  p^\m \tau +   f^\m_+ ( \sigma + \tau) + f^\m_- ( \sigma - \tau) $ 
subject to the closed string periodicity condition 
$x^\m (\tau,\s) = x^\m (\tau,\sigma + 2 \pi)$  is parametrized  by 
 constants,  
$f^\m_\pm (\s\pm \tau) = \sum_n 
  \big( a^{\m}_ {(\pm)n}  \cos[ { n (\s\pm \tau) }]   + b^{\m}_{ (\pm) n}  
  \sin[ {n (\s\pm \tau) }]\big),   $
which are constrained by the Virasoro conditions. 
Simple  explicit solutions representing semiclassical (coherent) 
states corresponding to particular quantum states  in the 
string spectrum have only 
 finite number of the Fourier modes excited. 
The  Virasoro  condition then   implies a 
relation  between the energy of the string $E=   T_0 \int d \s  \ \del_\tau  x^0 $ 
and its  linear momenta, spins,  oscillation numbers, etc. 
Some explicit  examples  are:

\noindent 
{\bf  Folded  string  rotating  on a  plane}:
\bea  
&&  x^0 = \k \tau \ , \ \ \ \ \ \ \ \ \ \ 
x_1+i x_2 = a \sin \s  \ e^{ i \tau}  \ , \la{fol}  \\ 
&& \la{fold} 
E = 2 \pi T_0 \k = \sqrt{ {\te{2\ov  \a'}}  J }\ , \ \ \ \ \ \ \ \   \ \  J=  {\te{  a^2 \ov 2\a'}} \ .   \eea
 \noindent 
  {\bf Spiky   string   rotating  on a  plane: }
 \bea
   &&  x^0 = \k \tau \ , \ \ \ \ \ \ \ \ 
x_1 + i x_2  =  \ha a \big[ e^{i m (\tau + \sigma ) }     +  m e^{i(   \tau- \sigma ) } \big] 
 \ ,\la{spi}  \\ 
&& E=  \sqrt{  {\te  {4m \ov (m+1) \a'} } J }   \ , \ \ \ \ \ \ \  \k =  a m  \ , \ \ \ \ \
J= {\te {a^2 m (m+1) \ov 4 \a'}}   \ . 
\eea
Here $m+1$ is the number of spikes, i.e. 
 $m=1$ is the case of the folded string.\foot{The relation between \rf{spi} and \rf{fol} for $m=1$ 
involves $\s \to \s + {\pi \ov2}$.}

\noindent  {\bf Circular  string   rotating  in two orthogonal planes of   $ R^4$: }
\bea 
&&x^0=\kappa\tau  , \ \ \ \ 
x_1+ i x_2 =  a \ e^{i ( \tau   +\sigma) } , \ \ \ \ \ \ \ 
  x_3+ i x_4 =  a\  e^{i  ( \tau -\sigma) }   \la{ff} \\
&&
E  = {\te  {\kappa \ov \a' } } = \sqrt{ {\textstyle{4\ov  \a'}}  J }\ , \ \ \ \ \ \ \ \ 
 \ \  J_1=J_2=J =  {\te{  a^2 \ov \a'}}  \ .   \la{yy} \eea
 Here $J_1= J_{12}, \ J_2 = J_{34}$ are  the values of the orbital momentum.

\noindent  {\bf  Circular string pulsating    in one plane}:
\bea \la{puu}
 && x^0 = \k \tau \ , \ \ \ \ \ \  x_1+i x_2 = a \sin \tau  \ e^{ i \s}  \ , \\ 
&& E= 2\pi T_0 \k =\sqrt{ {\te{2\ov  \a'}}  N  }  \ , \ \ \ \ \ \   N= {a^2\ov 2 \a'}\ .  \la{pul} \eea
Here $N$ is the oscillation number (an adiabatic invariant). 
This solution is formally not rigid but is very similar -- 
  the shape of the string remains circular, only  its radius changes with time. 
An  example of a non-rigid  solution is a   ``kinky string''  \ci{mac} for which 
 the string  has a shape of a quadrangle  at the initial moment in time,  then  shrinks to diagonal 
 due to  the tension, then expands back, etc.

\subsection{Circular rotating strings: rational solutions}

A simple  subclass of  ``rational'' 
solutions
 of the \adss equations \rf{ade},\rf{sse}  is 
found by assuming that $\Lambda, \td \Lambda =\const$  \ci{ft2,art}.
 In this case $Y_P$ and $X_M$ are given by simple trigonometric solutions of 
 the linear 2-d massive scalar  equation  and one is just to make sure that 
 the constant parameters are such that all the  constraints in \rf{ade}--\rf{peri} are satisfied.   
An example is  a circular string solution in $R_t \times S^5$  part of \adss 
 which is a direct analog of the circular 2-spin solution \rf{ff}  \ci{ft2}
 (see \rf{relx}) 
\bea
 \YY_0  =e^{i \kappa\tau}   , \ \ \ \ 
  \XX_1= {\te {a\ov \sqrt 2}}  \  e^{im(\tau+\sigma)}, \    \ \ \XX_2={\te {a\ov \sqrt 2}}  \ 
   e^{im(\tau-\sigma)}, ~~\ \  \ 
  \XX_3 =  \sqrt{ 1-  a^2 } \ , 
\la{sw}\\
 {\J}_1={\J}_2\equiv  \J = {m a^2\ov 2}= {\k^2 \ov 4m} , \ \ \ \ \ \ 
E=\sql \kappa = \sqrt{4m \sql J} \ . \la{eer} \eea 
Here $m$ is a winding number, 
$\td \Lambda= \k^2, \ 
\Lambda=0$, i.e.  the $S^5$  part of the solution is essentially the same as 
in flat space: the string rotates on $S^3$ of radius $a \leq 1 $ inside $S^5$ of radius 
1.  
 The semiclassical spin parameter $\J$ is bounded from above, i.e. 
the fast-string BMN-type   limit ($\J\to \infty$)   cannot be realised. 
Instead, there is a smooth small spin ($\J \to 0$)  or  ``small-string'' 
limit ($a \to 0$) in  which the Regge form of the energy is to be expected.   
Remarkably, the {exact}  expression for the classical string  energy 
  has  the same ``Regge'' 
form as in flat space \rf{yy} with $ {1 \ov \a'} \to \sql$.
This  solution is thus a  semiclassical analog \ci{rt} of a ``short''  quantum string for which the energy  should scale (for fixed charges) as 
 $E \sim  \sqrt{\sql}$  \ci{gkp2}. 
 The  solution \rf{sw}   has an obvious generalization to the case of the 3-rd non-zero spin 
in $S^5$ \ci{ft2}:  one needs to  consider a non-zero $ \XX_3 =  \sqrt{ 1-  a^2 } e^{i w' \tau} $. 

There is a  different  solution 
(with $\Lambda= w^2 - m^2$) 
describing a circular string with two equal  spins  moving on a  ``big'' $S^3\subset 
 S^5 $  \ci{ft2}
\bea
 \YY_0  =e^{i \kappa\tau}   , \ \ \ \ 
  \XX_1 ={\te{ 1 \ov \sqrt 2} }   e^{i(w\tau+m\sigma)}, \ \   \ \
   \XX_2= {\te {1 \ov \sqrt 2}}    e^{i(w\tau-m\sigma)}, ~~\ \  \ \XX_3=  0\ , 
\la{jk}\\
{\J}_1={\J}_2\equiv  \J =  \ha  w  , \ \ \ \ \ \ \k^2 = w^2 + m^2   , \ \ \ \ \ \ 
E=\sqrt{ (2J)^2 + \l m^2 }  \ . \la{eera} \eea 
The two solutions coincide when $a=  {1 }$ in \rf{sw} 
and $ w=m$ in \rf{jk}. 
This   solution  admits the   fast-string  limit  in which ($\J \gg 1$) 
\be 
E= 2 J + { \l  m^2 \ov 4 J}  -  { \l^2  m^4 \ov 64 J^3 }  + O( {\l^3 \ov J^5}) \ , 
\la{jjj} \ee
but it does not have a small-string limit as here the radius of the string is 
always 1: even though  $\J$ may become 
 small, the   energy does not go to zero  due to string winding around big circle of 
 $S^5$. In contrast to \rf{sw}, this solution is 
 unstable under small perturbations \ci{ft2,ft4}.

There is   another counterpart of   the flat-space 
solution \rf{ff} in \adss when the 
circular   string  rotates solely  in  $AdS_5$ \ci{ft2,art}
(here we  choose the  winding number to be  $m=1$)
\bea \la{spii}
\YY_0 = \sqrt{1 + 2 r^2}  \  e^{i \k \tau} \ , \ \ \ \ \ 
\YY_1  = r \  e^{i ( w \tau +   \s) } \ , \ \ \ \ \ 
\YY_2 = r \  e^{i ( w \tau -   \s) } \   . 
\eea 
Here   $ r=\sinh \r_0 =  { 1 \ov 2}\k  , \   w^2 = \k^2 + 1  $  and   
 the energy $E=\sqrt{\lambda}{\rm E}$. The two equal 
 spins $S_1=S_2 = \ha S=   \sql {\rm S}$  and the energy are related by  the parametric equations 
${\rm S} = { 1 \ov 4} \kappa^2\sqrt{\kappa^2 + 1 }  ,$  \ 
${\rm E}=\kappa + \ha \kappa^3 .$
This solution again admits a ``small-string''  limit ($\S\to 0$)
in  which it  represents a small  circular string   rotating around its c.o.m. 
 in the two orthogonal 
planes   in the central ($\rho\approx 0$ or ``near-flat'', see  \rf{ads}) 
  region of $AdS_5$. 
In the small spin limit $ {\rm S} \ll 1$  \ci{rt} 
\be\la{clas}
E=\sqrt{4\sqrt{\lambda}S}\ \Big[1+\frac{S}{\sqrt{\lambda}}
   - \frac{3S^2}{2\lambda}+ O({ S^3 \ov \l^{3/2}} )\Big] \ . 
\ee
Here in contrast to the $J_1=J_2$ solution \rf{sw} 
the classical  energy 
contains  non-trivial ``curvature''   corrections 
which 
 modify  the leading-order flat-space Regge behavior.
 In the opposite large spin  limit $ {\rm S} \gg 1$  we get \ci{ft2,art,ptt2}
\be \la{larh}
E= 2S + {3\ov 4} (  4 \l S)^{1/3} + O(S^{-1/3}) \ . \ee 
Yet another \adss  counterpart  of the flat-space 
solution \rf{ff} is found by having a circular string rotating both in $AdS_5$ 
and in $S^5$ (we choose  again the winding numbers in $\s$ to be 1) 
 \bea \la{mm}
\YY_0  = \sqrt{1+r^2}  \  e^{i \k \tau} \ , \ \ \ \ 
\YY_1 = r \  e^{i ( w \tau +  \s) } \ , \ \ \  \
 \XX_1  = a \  e^{i ( \tau -  \s) } \   , \ \  \ \ \XX_2  = \sqrt{1 -  a^2}      
\eea
Here $w^2 = \k^2 +1$ and $r=\sinh \rho_0$ and $a=\sin\g_0 $   
determine the size 
of the string in $AdS_5$ and $S^5$ respectively  (cf. \rf{ads},\rf{ses}). 
The conformal gauge    conditions \rf{cv} imply 
$   (1 + r^2 ) \k^2 = r^2 ( w^2 +1) + 2 a^2 , \ \ 
     r^2 w = a^2    $
 and thus for this solution one  has $\S=r^2w =\J=a^2 \leq 1 $, i.e.
  $S=J \leq \sql $. Also,    
$\E= (1+r^2)  \k =   \k +  { \S \k \ov \sqrt{ \k^2 +1 } } $, 
where $\k$ satisfies 
$ \k^2 = { 2 \S  \ov \sqrt{ \k^2 +1} } +   2 \S$
which is readily solved. In the small S limit one finds    (cf. \rf{clas}) 
\be 
E = \sqrt{4\sqrt{\lambda}S}\ \Big[1+\frac{S}{2 \sqrt{\lambda}}
   - \frac{5S^2}{8\lambda}+ O({ S^3 \ov \l^{3/2}} )\Big] \ . 
 \la{trt}\ee
In the small-size  or  $\S=\J \to 0$ limit (when $w\to 1, \ r \to a\to 0$) this 
solution reduces to the flat-space  one \rf{ff}  with the energy taking 
  the   form   \rf{yy}. 
 
At  the $\S=\J=1$ point  (where $a=1, \ \k= \sqrt 3, \ w=2, \ r = \sqrt 2$) 
 this ``small-string'' $S=J$  solution coincides 
with the   ``large-string'' $S=J$ 
solution discussed in  \ci{art,ptt1}
\bea
\YY_0 = \sqrt{1+r^2}  \  e^{i \k \tau}  , \ \ \ \ \ \  
\YY_1 = r \  e^{i ( w \tau +  \s) }  ,  \ \ \ \ \ \  \ 
  \XX_1   = e^{i ( \omega \tau -  \s) } , \la{iop}\\ 
w^2 = \k^2 +1 ,\ \  \ \ \ \       \S= r^2 w = \omega = \J   \ . \eea
  Then   
  $\E= \k+  { \S \k \ov \sqrt{ \k^2 +1 } }$, where $\k(\S)$ satisfies 
$ \k^2 = { 2 \S  \ov \sqrt{ \k^2 +1} } +    \S^2 +1$.
The  cubic equation for $\k^2$ admits two real  solutions 
$
\k^{(1,2)} = \sqrt{1 + \ha \S^2 \pm  \ha \S \sqrt{8 + \S^2}} .$
The first solution  is defined for
 any $\S \geq -1$ and the  corresponding energy  \ci{rt}
\be 
  \E= \sqrt{ 1 + \ha \S^2 + \ha  \sqrt{ 8 +  \S^2} } \  {\te { 
 \Big[ 1 + {  \S \ov   \sqrt{ 2 + \ha \S^2  + \ha  \sqrt{ 8 +  \S^2} }  }   \Big]}}
 \la{yu} \ee
 admits a  regular  large $\S$ expansion as in \rf{jjj}  \ci{art,ptt1}:  
 \be \la{swd} 
 E =  2 S + {\l  \ov S} - { 5 \l^2 \ov 4 S^3} +   O( { \l^3 \ov S^5}) \  .
\ee
In  the small $\S$ expansion we get 
$  E= \sql  + \sqrt 2\  S + { S^2 \ov 4\sql }  + ...   $, 
i.e. 
this   solution   does not have the flat-space
 Regge asymptotics; this is not surprising 
since here the string is wrapped   on  a big circle of $S^5$  and its tension    gives a large 
contribution to the energy even for small spin. 


The above examples illustrate possible  patterns of behaviour of the 
classical string energy on the string tension and conserved spins  in different limits.
They  should be reproducible  from the exact  results for the string spectrum in 
 appropriate semiclassical string limits.

\subsection{Rigid string ansatz: reduction to   1-d  Neumann   system}

The above examples of solutions in \adss  are special cases of a 
rigid string ansatz  for which the shape of the string does not change with time 
$\tau$ or the AdS time $t$.
Making such an  ansatz  and substituting it into the equations 
\rf{ade},\rf{sse}  one finds that they can be obtained  from a 1-d integrable 
action describing an oscillator on a sphere  -- the Neumann model \ci{afrt,art,ft2}. 
Along with  the integrability of the equations describing geodesics in \adss 
this reduction of the \adss  string sigma model to  an integrable 1-d 
system is a simple illustration   of the  {\it integrability}  of this 2-d theory. 

The general solution of the resulting equations  can then be  written
 in terms of hyperelliptic (genus 2 surface)  functions, with the  rational solutions
  discussed above and the  elliptic  solutions  described below in the next section
    being the  important special cases. 
The general rigid string ansatz may be written as (see \rf{relx})
\be \la{ge}
\YY_r= \zz_r(\s )\ e^{i \w_r\tau }\  \  \ \ \  (r=0,1,2) \  ; \ \ \ \ \ \ \ \ 
 \ \ \ 
\XX_i=z_i(\sigma)\ e^{iw_i\tau}\  \ \ \ (i=1,2,3) 
\ee
Here $\w_{1,2}$ and $w_i$ are rotation frequencies and $\zz_r$ and $z_i$ 
(which are,  in general,  complex) satisfy 
\be\la{dos} 
\zz_r = \rr_r e^{i \b_r} \ , \ \ \ \ \ \ 
\eta^{rs}\rr_r\rr_s = -1\ , \ \ \ \ \ \ 
z_i =  r_i   e^{ i \a_i}\ , \ \  \ \ \ r^i  r_i =1 \ , 
\\   \la{hi}
\rr_r (\s + 2 \pi)= \rr_r (\s) \ , \ \ \ \ \ \ 
\b_r (\s + 2 \pi)= \b_r (\s)  + 2 \pi k_r \ , \\ 
  r_i (\s + 2 \pi) = r_i (\s ) \ , 
 \ \ \ \ \ \
\a_i(\s+ 2 \pi ) =  \a_i(\s)  + 2 \pi  m_i   \  .\la{opi}  \ee 
 Here $\eta_{rs}=(-1,1,1)$, \  $k_r$ and $m_i$ (which are the ``winding numbers'' 
 for the corresponding isometric angles in \rf{relx}) 
 are integers.
We  assume that $\beta_0=0,$ $  k_0=0,$ $  \omega_0 \equiv \kappa$.
The corresponding Cartan charges are 
(cf. \rf{cha},\rf{aas},\rf{jij})\foot{Here $E= S_0$. All  other components 
of the conserved angular momentum tensors 
in \rf{cha}  vanish
  automatically    if all the frequencies are 
different \ci{afrt}, but their vanishing should be checked 
if 2 of the 3  frequencies are equal.}   
$
\S_r= \w_r \int_0^{2\pi } {d\sigma\over 2\pi }\ 
\rr_r^2(\sigma )  , \ \ $ $ 
\J_i =   w_i
\int_0^{2\pi}\frac{d\sigma}{2\pi}\ 
 r_i^2(\sigma).$
The equations for the  remaining ``dynamical'' variables $\rr_r$ and $r_i$ can 
be derived from  the following 1-d ``mechanical''  Lagrangian 
\be 
L =  \eta^{rs} (\zz_r'{\zz'^*_s}-\w_r^2  \zz_r\zz_s^* )- \tilde \Lambda (\eta^{rs} \zz_r\zz_s^*   + 1)\  + \  
z'_i z'^*_i - w_i^2 z_i z_i^*
+  \Lambda ( z_i z_i^*-1)  \ . \la{nm}
\ee
The trajectory of this effective ``particle''
belonging  to a product 
of a 2-hyperboloid ($\rr_r$)  and 
  2-sphere ($r_i$)
 gives the  profile of the string.
The angular parts 
of $\zz_r$  and $z_i$ can be  easily separated leading to an effective Lagrangian 
for a particle on a constant curvature surface  with an  ``$r^2 + r^{-2}$'' potential
or to a special case of a 1-d integrable   Neumann system -- 
the Neumann-Rosochatius system \ci{art}. 
The corresponding 2+2 integrals 
of motion can be explicitly written down  \ci{afrt,art}.
The resulting   solutions represent, in
particular,  folded or circular   bended wound  rotating rigid strings.

 For example, such  closed string solutions in $AdS^5$   will
  be  parametrised by  the frequencies  $\om_0=\k, \om_i=(\om_1,\om_2)$
  as well  by  two integrals of motion $b_k$. \   $(\om_i,b_k)$
 may be viewed as independent  coordinates on the  moduli space of  these solitons.
The closed string periodicity condition
 implies that the  solutions will be classified
by  two    integer ``winding numbers''   $n_i$  related to  $\om_r$ and $b_i$.
 In general, the energy
$E$ will be a function not only of $S_1,S_2$ but also of  $n_i$.
Depending on the values of these parameters  the string's   shape may be of the 
two  types:
(i)
 { ``folded''}, i.e. having  a shape  of an interval,
 or  (ii)  { ``circular''},  i.e. having  a shape  of a circle.
 A folded string may be  straight as in the one-spin  case  \ci{gkp}
 or bent \ci{afrt,tt2}.
 A ``circular'' string  may  be  a round circle as in  \ci{ft2}  or
 may  have a more general  ``bent circle'' shape.
Some of such  solutions will be discussed explicitly below.

\section{Spinning  rigid strings  in \texorpdfstring{$AdS_5\times S^5$}{AdS5xS5}: \texorpdfstring{\\}{} elliptic solutions}

In this section we shall   consider an  important  example 
of a non-trivial  rigid string solution 
describing a folded spinning string in $AdS_3$ part of $AdS_5$ \ci{dev,gkp}.
We shall  then discuss some of its generalizations and  
similar solutions described in terms of  elliptic functions. 

\def \om {\omega}


\subsection{Folded spinning string  in \texorpdfstring{$AdS_3$}{AdS3}}

Let us consider a rigid string moving in $AdS_3$ part 
$ds^2= -\cosh^2 \rho\ dt^2 + d \rho^2 + \sinh^2 \rho\ d \phi^2$
of $AdS_5$ \rf{ads}, i.e.
$\YY_0 = \cosh \rho(\sigma)\ e^{ i \kappa \tau},$  \ 
$\YY_1 = \sinh \rho (\s) \ e^{ i \om \tau} $, 
or 
\be
\label{sol}
t= \kappa \,\tau,~~ \quad \phi= \om \,\tau, ~~\quad \rho=\rho(\sigma) =\r(\s+2\pi)  \ . 
\ee
This  ansatz satisfies the equations for $t$ and $\phi$ while for $\rho$ we get 1-d 
sinh-Gordon equation 
$
\rho'' = \ha (\k^2 - \om^2) \sinh (2 \rho)$. Its  first  integral 
satisfying the Virasoro  condition \rf{cv} 
leads to the following solution 
\be\label{eom}
\rho'^2 = \kappa^2\cosh^2\rho-\om^2\sinh^2\rho, \\\label{rhop}
\sinh\rho  (\sigma) = \frac{k}{\sqrt{1-k^2}}\, \cn(\omega\,\s+\KK\,|\, k^2) \ , \ \ \ \ \ \ \ \ \ 
k\equiv {\kappa\ov \omega} \ . 
\ea
Here we assumed that $\rho(0) =0$;   
cn is the  standard elliptic function  and 
$\KK\equiv \KK (k^2)= \int_0^{\pi/2} du (1 - k^2 \sin^2 u)^{-1/2}$ 
 is the { complete elliptic integral of the first kind}.
This  solution  describes a folded closed 
 string rotating around its center of mass and  generalizes the flat-space 
 solution \rf{fol} (for  $\s \to  0$ we get $ \sinh \r \to 
      a \sin \s $, \ $a= \frac{k}{\sqrt{1-k^2}} $). 
In \rf{rhop}   $\sigma $ varies from 0 to $\pi \ov 2$ with $\rho$  changing 
 from $0$ to its maximal  value $\rho_0$, \ 
$
\coth \rho_0 = \frac{\om}{\k}=  k^{-1}$. 
The full ($2 \pi$ periodic)
  folded closed string solution
  is found by   gluing  together four such functions $\rho(\sigma)$ 
   on $\pi \ov 2$ intervals
 to  cover the full $0\leq \sigma \leq 2 \pi$ interval. 
The periodicity condition 
$ 2\pi=\int_0^{2\pi}d\s=4\int_0^{\rho_0}\frac{d\r}{\sqrt{\k^2\,\cosh^2\r-\om^2\,\sinh^2\r}}
$
implies a relation  between  the parameters $\k$ and $\om$,  
 i.e. 
$ \k=\frac{2\,k}{\pi}\,\KK  ,\ \ ~~\om=\frac{2}{\pi}\,\KK .$
 The 
 classical energy $E=\sql \E$  and the spin $S=\sql \S$ 
 are expressed in terms of
 the complete elliptic integrals 
   $\KK=\KK(k^2)$ and $\EE=\EE(k^2)=\int_0^{\pi/2} du (1 - k^2 \sin^2 u)^{1/2}$  
 \ba\label{Ec}
{\E}= \frac{2}{\pi}\,\frac{k}{1-k^2}\,\EE \ ,  \ \ \ \ \ \ \ \ \ \ \ 
{\S}= \frac{2}{\pi}\,\big(\frac{1}{1-k^2}\,\EE-\KK \big) \ .
 \ea
Solving for $k$ gives the relation $\E=\E(\S)$. 
The expression for  $\E(\S)$  can be
easily  found  in the two limiting cases: 
(i)  large spin or long string  limit: \  $\r_0\to\infty$, i.e.  $k\to 1$, and 
(ii) small spin or short string limit: \    $\r_0\to0$, i.e.   $k\to0$. 
In the  first  limit   the string's  ends are 
close to the boundary of $AdS_5$  and  one obtains 
 \cite{gkp,ft1,bftt}
\ba
{E}= S+ { \sql \ov \pi} \big[  {\ln ({\te {8\pi\ov \sql}} S )  -1}  \big] 
 + { \l \ov 2 \pi^2} \frac{\ln ( {\te {8\pi \ov \sql}} S)  -1}{ S}+  O ( { \ln ^2 S \ov S^2} )   \ , \ \ \ \ \ \ \ \ 
\S\gg1 \ .  \label{Ecl} 
\ea
The coefficient of the  $\ln S$ 
 term \ci{gkp} is governed by the strong-coupling limit of the 
 so-called ``scaling function" (cusp anomaly) and
 the subleading terms  can be shown to obey non-trivial
  reciprocity relations~\cite{bk,bftt} (see  \ci{IIId}). 
The leading $\S$ term in \rf{Ecl} \ci{dev}  may be interpreted as  being   due to the
  fold points of the string  moving (in the strict $\S=\infty$ limit)
along  null lines  at the boundary  while  the 
$\ln \S$ term \ci{gkp}  is due to the stretching of the string 
(this term is  string length times its 
tension). Indeed,  in the large spin limit  or $\k,\om \gg 1$   the solution \rf{rhop}
with  $\s \in ( 0, {\pi \ov 2})$ simplifies  to \ci{ft1,ftt}\foot{This is readily 
seen directly from \rf{eom} in the limit when $\k \to \om$.} 
\be \la{asi}
t= \kappa \,\tau,~~ \quad \phi= \om \,\tau, ~~\quad 
\rho = \k \s  \ ,\ \  \ \ \ \   \ \k=\omega \   \gg \ 1   \ .   \ee
This very simple form  of the asymptotic large spin solution allows one to compute 
quantum 1-loop \ci{ft1} and 2-loop \ci{rtt}  corrections to the energy (see \ci{IIc,IId}).

Let us   mention  also   that the 
 asymptotic solution \rf{asi} with $\k \to \infty$   describing  infinite  string  with folds 
 reaching  the AdS boundary  and capturing  the coefficient 
 of the $\ln S$ term in $E-S$  \rf{Ecl}  is closely related to the ``null cusp''  open string solution 
 \ci{kruk} 
 describing an  open string  (euclidean) 
 world surface  ending on  the two orthogonal  null lines at the boundary of 
 $AdS_5$ in Poincare coordinates,  $z= \sqrt{2 x^+ x^-} , \ \     x^\pm = x_0 \pm x_1$  (see \rf{aa}).
 In  the conformal gauge 
 \be \la{cus}
 z= \sqrt 2\ e^{ \sqrt 2 \tau} \ , \ \ \ \ \ \ 
 x^+ = e^{ \sqrt 2 ( \tau  + \sigma ) }  \ , \ \ \ \ \ 
 x^- = e^{ \sqrt 2 ( \tau  - \sigma ) }  \  . \ee
This solution written  in the  embedding coordinates \rf{eme}  is then equivalent to  
\rf{asi} after a euclidean continuation ($\tau \to i \tau$) and an $SO(2,4)$  coordinate transformation 
\ci{krutt}. This explains (from strong-coupling or semiclassical string perspective) 
why  the coefficient  of the $\ln S$ term  in \rf{Ecl}  can be interpreted as a cusp anomalous  
dimension  (a dimension of a Wilson loop defined by null cusp,  
see also \ci{IIId,IVb}). 

In the   small spin or ``short string''  limit, when   the string is rotating  
 in the  central ($\rho=0$) region of $AdS_3$  we get 
 the same  flat-space \rf{fold} Regge type asymptotics ~\cite{gkp,ft1,tt} as 
 in the  circular string cases in \rf{clas},\rf{trt}   
\be
 E=\sqrt{2 \sql S}\ \Big[1 + { 3 S\ov 8\sql}  +  O(S^2)  \Big] \ , \ \ \ \ \ \ \ \ \ \ \ \ 
\S\ll1 . 
\label{Ecc}
\ee

\subsection{Some  generalizations and similar solutions}

The above  $AdS_3$ solution is  special  having minimal  energy for given spin. It has several
  generalizations. 
One may consider a similar  solution 
 of  circular shape  with several spikes \ci{krus}   that is  the analog of 
the spiky string  in flat space \rf{spi}.\foot{The spiky string 
 is described (in conformal gauge) by a generalization of the ansatz 
in \rf{ge}  discussed  below.} 
For the spiky string in AdS  the large spin limit of the energy is (cf. \rf{Ecl})

\noindent
$E= S +  { \sql n  \ov 2 \pi} \big(  \ln {16 \pi S \ov \sql n  } -1 
+ \ln \sin { \pi \ov n} \big)  + ...\
, $ 
where $n$ is the number of spikes
($n=2$ is  the folded string case). 
 The  large-spin asymptotic solution 
consists of $n$ segments  each of which is conformally  equivalent to the limit 
\rf{asi} of the  folded string \ci{krust}. 

One  may also  find  similar rigid string   solutions with   $\ln S$ scaling
 of $E-S$  at large spin  with  two non-zero 
 spins $S_1, S_2$, i.e. moving in the whole $AdS_5$  \ci{ft2,afrt,kl,r,tt2}  
subject to the rigid string ansatz \rf{ge}, i.e. 
$
 t = \kappa \tau  , \ \    \r= \r(\s) , \ \   \theta=\theta (\s)  , \ \ 
 \phi_1 = \om_1 \tau  , \ \   \phi_2 = \om_2 \tau $. 
 The  simplest  circular solution of that  type  is a  round  string \ci{ft2}  with
 $\r=\r_0=\const , \ \theta= {\pi \ov 4} , \ \ \om_1=\om_2$
 and thus with $S_1=S_2$  already discussed above in \rf{spii}-\rf{larh}.
 It does not, however, represent a state with a
 minimal energy for given values of the spins.
 To get a stable  lower-energy solution with $S_1=S_2$ one is to relax
 the $\r=\const$ condition,   allowing the string to develop, in the large spin limit,
  long arcs  stretching to infinity (i.e. to  the boundary of $AdS_5$)
  and carrying most of the energy. Then 
 for a  particular $S_1=S_2= S $  string of circular shape with with 3 cusps 
 described by an elliptic function limit of a  general hyperelliptic 
 solution of the Neumann model \rf{nm} 
  one  finds  for its energy \ci{kl,tt2}: \ 
$E= 2S +  \frac{3}{2} \times  { \sql \ov \pi} \ln S +...$.
Similar open-string  solutions were discussed  in \ci{Irrgang:2009uj}. 

Another  important generalization of the folded 
spinning string in $AdS_3$   is found by adding an 
angular  momentum  $J$ in $S^5$,  i.e. by  assuming  in addition to 
\rf{sol} that the string orbits a  big circle in $S^5$, $\vp = \nu \tau$ \ci{ft1}.
The $AdS_5$ and $S^5$ parameters are coupled via the Virasoro constraint 
\rf{eom} which is  modified to 
$\r'^2 = 
(\k^2-\nu^2) \cosh^2 \r - (\om^2-\nu^2)  \sinh^2 \r $  so that 
the relations \rf{rhop}--\rf{Ec} have straightforward generalizations. 
The resulting expression for the energy $E= \sql \E( \S,\J)$  (with $\J=\nu$) 
 can be expanded 
in several limits. 
In the short string limit with $\J \ll 1, \ \S \ll 1$ one finds 
\ci{ft1}
\be \la{pxx}
E= \sqrt{J^2 + 2 \sql S } + ... \ . \ee
This limit   probes  the $\r\approx 0$ region  of 
$AdS_5$    where  the energy spectrum
should thus be as in flat space, i.e. should be  just a  relativistic  expression
for the  energy of a string 
 moving  
with momentum $\J$   and rotating around its  
 c.o.m. with spin $S$, i.e.   $E^2 - J^2 = 2 \sql S + ...$. 
If the  boost energy  is smaller than the rotational  one,
 $ \J^2 \ll \S$, then
$E \approx \sqrt{2\sql S}   +  O ( {J^2 \ov \sqrt S} ).$
For  strings with  $\J \gg 1$  and  $\J \ov  \S$=fixed  we get a  regular 
``fast-string''  expansion 
as in \rf{jjj},\rf{swd}, 
$
E=   J   + S   +  {\l   S \ov 2 J^2 } + ... $. 
In the limit when $S$ is large
 the string 
can become very long and its ends 
approach the boundary of $AdS_5$.
The
analog of the asymptotic solution \rf{asi} is 
\be \la{assi}
\rho = \mu \s, \ \ \ \ \ \k= \om, \ \ \ \ \
\mu^2 = \k^2 - \nu^2  \ , \ \ \ \ \ \ \ \k, \mu, \nu \gg 1 \ .  \ee
The spin $S$ and $\mu$  are related by   $\mu \approx 
{ 1 \ov \pi} \ln \S + ...$ so in the limit when $\k,\om,\mu,\nu$ are
 large  with  their ratios  fixed,  
i.e.  $ \S \gg 1$ with $\ell \equiv { \pi \J \ov \ln \S }=$fixed  we  get \ci{bgk,ftt,russ}
\be \la{hew}
E= S + {\te {  \sqrt{ J^2 +  { \l \ov \pi^2} \ln^2 { S } }}} + ...  
= S   +  {\sql\ov \pi}  f_0(\ell) \ln S + ... 
  \ , \ee 
where $ f_0(\ell) = \sqrt{ 1 + \ell^2} $. 
Again, the fast-string  expansion in the limit when  $\ln \S  \ll \J $  (i.e. $\ell\gg 1$) 
 gives 
a regular  series in $\l$ \ci{ft1}, 
$E=S +J+\frac{\lambda}{2 \pi^2 J}\, \ln^2 \frac{S}{J} +....$.
This solution has also a generalization to the case of winding along 
$S^1$ in $S^5$ \ci{grrtv,kum}.

There is also  an analog of the folded  spinning string 
in $S^5$ \ci{gkp}, where the string  is spinning on $S^2$  with its center 
at rest. The corresponding ansatz is $ X_1 + i X_2  = \sin \psi(s) \ e^{i w \tau}, \ 
\ X_3 = \cos \psi(s)$  where $\psi$ solves the 1-d sine-Gordon equation.  
 The   short string (small spin)  limit here  gives again the  flat-space 
Regge behaviour,

\noindent $E= \sqrt{2\sqrt{\lambda} J}\ 
\big(1+\frac{ J}{8\sqrt{\lambda}}+...\big)$. For large spin 
 $E = J + { 2} { \sql \ov \pi}   + O( J^{-1}) $. 

 There is  a $(J_1,J_2)$ generalization of this 
 solution discussed in \ci{ft3,bfst}. 
 The AdS spiky string of \ci{krus}  also admits a generalization to
  the case of non-zero $J$ or/and winding in $S^1 $ of $S^5$ 
 \ci{kruj}; in this case the  spikes are rounded up. 
 
Among other elliptic solutions let us mention also 
pulsating  strings in \adss  that generalize the flat space solution \rf{puu}
\ci{gkp,minahan,krtt,ptt2}; here  the  role of the spin 
is played by the adiabatic invariant -- the  oscillation number $N= { \sql \ov 2 \pi} 
\int d \theta p_\theta$. 
It is interesting to compare the large/small spin expansions of the
 classical string energy
in the equations 
\rf{jjj},\rf{clas},\rf{larh},\rf{trt},\rf{swd} and   \rf{Ecl},\rf{Ecc} 
with what one finds for pulsating string solutions  in $AdS_3$ \ci{minahan,ptt2}
($\NN = { N \ov \sql}$) 
\ba
&&{E}= N + c_1 \sqrt{ \sql N} +  O(\NN^0)    \ , \ \ \ \ \ \ \ \ 
\NN \gg1 \  , \ \ \ \ \  c_1 = 0.7622...  \label{hlo} \\
&&{E}= \sqrt{ 2 \sql N} \ \Big[1 + { 5 N\ov 8\sql}  +  O(\NN^2)  \Big] \ , \ \ \ \ \ \ 
\NN \ll1 \  , \la{kpy} 
\ea
and $R \times S^2$ \ci{minahan,krtt}
\ba
&&{E}= N +  { \l \ov 4 N} +   O(\NN^{-2})    \ , \ \ \ \ \ \ \ \ 
\NN \gg1 \  ,    \label{hloi} \\
&&{E}= \sqrt{ 2 \sql N} \ \Big[1 - {  N\ov 8\sql}  +  O(\NN^2)  \Big] \ , \ \ \ \ \ \ 
\NN \ll1 \  . \la{kuy} 
\ea

\subsection{Spiky strings  and giant magnons in \texorpdfstring{$S^5$}{S5}}

An important class of rigid strings  that  are described 
by a slight generalization of the  ansatz in \rf{ge}
 are  strings with spikes
 \cite{krus,Ryang} and 
 (bound states of) 
giant magnons \cite{HM,Dorey,Dorey2} with several non-zero  angular momenta.
Both the spiky strings in $S^5$ 
 and  the giant magnons can be  described \ci{krt} by a
  generalization of the rigid string  ansatz \rf{ge} of \cite{afrt,art}. 
It is possible to show  that  the giant magnon solutions
are a particular limit of the spiky string solutions  and that 
 a giant magnon  with two angular momenta can be interpreted
 as a superposition of two magnons moving with the same speed. 
Consider strings moving  in $R_t \times S^5$ part of \adss and described  by  
the following   generalization of  the rigid string ansatz in \rf{ge} \ci{krt}
\be \la{jm}
t= \k \tau \ , \ \ \ \ \ \ 
\  \XX_i = z_i(\xi)\ e^{iw_i\tau}\ , \ \ \ \ \ \ \ \ 
 \ \ \   \ \ \xi\equiv  \sigma +  b \tau \ , \ee
where $ z_i= r_i e^{i \a_i}   , \ \ 
  z_i (\xi+2\pi )=z_i(\xi) $. 
Here   $b $ is a new parameter.
The 1-d   mechanical  system for the functions 
$z_i$ that follows 
from 
 \rf{sse}
  is an integrable model:  a  generalization  of the  Neumann-Rosochatius  one 
where a  particle  on a sphere is coupled also to a constant magnetic field.  
This ansatz   describes the $S^5$ analog of the 
$AdS_5$ spiky string of \ci{krus}  with   extra  
angular momenta \ci{krt}.  The spiky string is built out of several arcs; 
in the  limit  when $J_1 \to \infty$ with $E-J_1$=finite 
the single arc  is  the giant magnon  of 
\ci{HM} with an  extra momentum  $J_2$ \ci{Dorey2} 
(see also \cite{AFZ,MTT}). 
In this limit   $\kappa \to \infty$ and  it is natural to rescale 
 $\xi$ so that it takes values on an infinite line (a 
 single arc is  an open rather than a closed string).  Then  
 \be
E-J_1  = {\te \sqrt{J_2^2 + \frac{\l}{\pi^2} \sin^2 \frac{p}{2}} } \ ,  
\label{twoJJ}
\ee 
where $p$ is related to the length of the arc and 
 may be  interpreted as a  momentum  of the giant magnon \ci{HM}. 
The giant magnon 
may be  viewed as a strong-coupling ``image'' of the 
elementary spin-chain magnon on the gauge-theory side. 

One may also  find  a  generalization  of the giant magnon  with two finite angular
 momenta $J_2,J_3$ \ci{krt}. 
 A single-spin 
  folded string  in $S^2$  \cite{gkp}
 in the limit when 
the  folds  approach the equator   can be interpreted \ci{HM} as a
superposition of two magnons with $p=\pi$ and $J_2=0$. A generalization 
 to the case of $J_2,J_3 \not=0$ is \ 
$E-J_1={\te  \sqrt {J_2^2+  {\lambda\over\pi^2 }} + \sqrt{J_3^2+ {    \lambda\over \pi^2 }}}.$
When  $J_2=J_3=0$, one recovers the expression for the energy of two giant magnons
with $p=\pi$, i.e.  $ E-J_1= 2{ \sqrt {\lambda}\over \pi} $ 
or 
 the leading term in the folded spinning string energy in the limit $J_1 \to \infty$.
Spiky strings  with several spins were discussed also in \ci{sprad,BR,Ryang:2006yq}. 

Let us mention also  some  related   rigid string  elliptic solutions.  
A   ``helical'' string  solution  interpolating between  the 
 folded or circular spinning string  and the 
 giant magnon  with spin was constructed  in 
 \cite{Okamura:2006zv}. 
Refs.   \ci{Ishizeki:2007we,Ishizeki:2007kh} found 
an ``inverted'' single-spike  string wrapping the equator 
of $S^2$  in $S^5$ (see also \cite{Abbott:2008yp}).
Ref. \cite{Hayashi:2007bq} (see also \ci{Okamura:2008jm} for a review) discussed 
a general family of ``helical'' string solutions in $R_t \times S^3$  
(which are most general  elliptic  solutions on $R_t \times S^3$)
interpolating  between pulsating 
 and single-spike strings which was  obtained from the helical 
string of  \cite{Okamura:2006zv} by interchanging $\tau$ and $\sigma$ 
in $S^3$ coordinates (this  maps  a string
 with large spin into a pulsating string  with large winding number).

\subsection{Other approaches to constructing solutions}

The integrability of the sigma model  equations \rf{ade},\rf{sse} 
implies that one is  able to  construct large relevant class  of 
solutions -- ``finite gap''  solutions in terms of theta-functions \ci{dooor}. 
Also one can  construct
 new  non-trivial solutions 
from given ones using ``dressing'' \ci{dr} or  B\"acklund transformations \ci{ba}.
 Using  the dressing method  one  may generate non-trivial solutions from simple ones, e.g.,  
  non-rigid or  non-stationary (scattering) solutions from  rigid string ones. Examples are   
  scattering and bound states of  giant magnons with several spins and arbitrary momenta 
 \ci{sprad,Kalousios:2006xy} or the   single-spike solution of \ci{Ishizeki:2007we} 
from a static wrapped string  and solutions with multiple  spikes 
 describing their scattering  \cite{Ishizeki:2007kh}.
  Similar methods can be applied also in the  open-string (Wilson loop) setting 
\ci{AM} to find generalizations of the null cusp solution \rf{cus} \ci{Jevicki:2007pk}. 
 

An alternative approach to constructing  explicit \adss  string solutions of 
\rf{ade}--\rf{cv} 
is based on the Pohlmeyer  reduction \ci{pohl,barb,deve,t2,HM,Dorey2,Okamura:2006zv,
poll,mik,mir,je}.  
The basic idea is  to solve the Virasoro  conditions \rf{cv} explicitly 
by introducing, instead of the string coordinates $(Y_P,X_M)$,  a new  set of 
``current''-type variables. Then \rf{ade}--\rf{cv}  become  equivalent 
to a generalized sine-Gordon (non-abelian Toda)  2-d integrable system.
Given a solitonic  solution of this system   one can then reconstruct 
the  corresponding string solution by solving linear equations  for $(Y_P,X_M)$ 
with $\tilde \Lambda$ and $\Lambda$ in \rf{ade},\rf{sse}  being   given 
 functions of $(\tau,\sigma)$. For example, in the case of a string on $R_t \times S^2$ 
 one may set $t =\kappa \tau$  and then the three 3-vectors 
 $X_i, \del_+ X_i, \del_-  X_i$  ($i=1,2,3$)  will  have  only one non-trivial scalar product
 $\del_+ X_i\del_- X_i \equiv  \k^2 \cos 2\alpha= - \Lambda$.  
 The remaining dynamical equation   takes the SG  form:
 $\del_+ \del_- \alpha   + {\k^2 \ov 2} \sin 2 \alpha =0$. 
 The Pohlmeyer-reduced model  for a string   on $R_t \times S^3$ is  the 
 complex SG  model,  while strings moving in $AdS_5$ are related to generalized sinh-Gordon-type  models.
   The giant magnon  corresponds to the  SG soliton \ci{HM}
 while  its $J_2\not=0$ generalization -- to  charged soliton  of the complex SG model \ci{Dorey2}. 
  Various  examples of solutions (multi giant magnons, spikes, etc.) 
  obtained using this method   can be found in  
 \ci{Okamura:2006zv,mir,je,Aniceto:2008pc}. 
 The approach based on the Pohlmeyer   reduction was recently applied also  to constructing 
 open-string surfaces  ending on null  segments which  generalize the null cusp solution 
  \ci{open1,open2}.

 
\phantomsection



\end{document}